\documentclass[twocolumn,showpacs,preprintnumbers,amsmath,amssymb,superscriptaddress,prl]{revtex4}

\usepackage{graphicx}
\usepackage{dcolumn}
\usepackage{bm}

\newcommand{\unit}[1]{\,\mathrm{#1}}

\begin{document}


\title{Directed percolation criticality in turbulent liquid crystals}

\author{Kazumasa A. Takeuchi}
 \email{kazumasa@daisy.phys.s.u-tokyo.ac.jp}
\author{Masafumi Kuroda}
\affiliation{
Department of Physics, The University of Tokyo, 7-3-1
  Hongo, Bunkyo-ku, Tokyo, 113-0033, Japan
}%

\author{Hugues Chat\'e}
\affiliation{
Service de Physique de l'\'Etat Condens\'e, CEA-Saclay, 91191 Gif-sur-Yvette, France
}%

\author{Masaki Sano}
 \email{sano@phys.s.u-tokyo.ac.jp}
\affiliation{
Department of Physics, The University of Tokyo, 7-3-1
  Hongo, Bunkyo-ku, Tokyo, 113-0033, Japan
}%

\date{\today}

\begin{abstract}
We experimentally investigate the critical behavior of a 
phase transition
between two topologically different turbulent states
of electrohydrodynamic convection in nematic liquid crystals.
The statistical properties of the observed
spatiotemporal intermittency regimes 
are carefully determined, yielding a complete set of
static critical exponents 
in full agreement with those defining the directed
percolation class in (2+1) dimensions.
This constitutes the first clear and comprehensive experimental evidence
of an absorbing phase transition in this prominent non-equilibrium 
universality class.
\end{abstract}

\maketitle

Transitions into an absorbing state,
 i.e. a state from which a system can never escape,
 arise from simple mechanisms expected to be widespread in nature.
Examples abound in a wide variety of situations
 in physics and beyond \cite{Hinrichsen-AdvPhys2000},
 ranging from spreading processes like epidemics and forest fires,
 to spatiotemporal chaos, catalytic reactions,
 and calcium dynamics in cells, etc.
Moreover, a host of problems such as
 synchronization \cite{Ahlers_Pikovsky-PRL2002},
 self-organized criticality \cite{Dickman_etal-BrazJPhys2000}, and
 wetting \cite{Ginelli_etal-PRE2003},
 can be mapped onto them.
Having no equilibrium counterparts, absorbing phase 
transitions are central in the ongoing search for the relevant ingredients
determining universality out of equilibrium.

Over the past 25 years or so,
it has been well established both in theory and in simulations that
the vast majority of absorbing phase transitions
share the same critical behavior,
constituting the so-called directed percolation (DP) universality class
\cite{Hinrichsen-AdvPhys2000}. 
This is not surprising since the DP class
corresponds to the simplest case of a single effective absorbing state
in the absence of any extra symmetry or conservation law,
 as conjectured by Janssen and Grassberger
 \cite{Janssen-ZPhysB1981,Grassberger-ZPhysB1982}
 and demonstrated by hundreds of numerical models
 \cite{Hinrichsen-AdvPhys2000}.

However, the situation is quite different at the experimental level.
Searching for evidence of DP critical behavior,
a string of experiments have been performed
\cite{DPexperiments},
but the results remained unsatisfactory:
they could not yield a complete set of critical exponent values in 
agreement with those defining the DP class (Table S1 \cite{EPAPS}).
In view of this state of affairs,
the importance of finding at least
one fully convincing experimental realization of DP-class scaling
has been stressed
\cite{DPopenproblem}.
The main difficulty probably stems from the fact that
 one must exclude long-range interactions \cite{Hinrichsen-AdvPhys2000},
 work with macroscopic systems to tame quenched disorder
 \cite{Hooyberghs_etal-PRL2003},
 and study them over long enough scales so that the critical behavior 
 is unambiguously measured.
We have overcome these difficulties and report here
 on a firm experimental observation of DP criticality.

\begin{figure*}[t]
 \begin{center}
  \includegraphics[width=171.427 mm,clip]{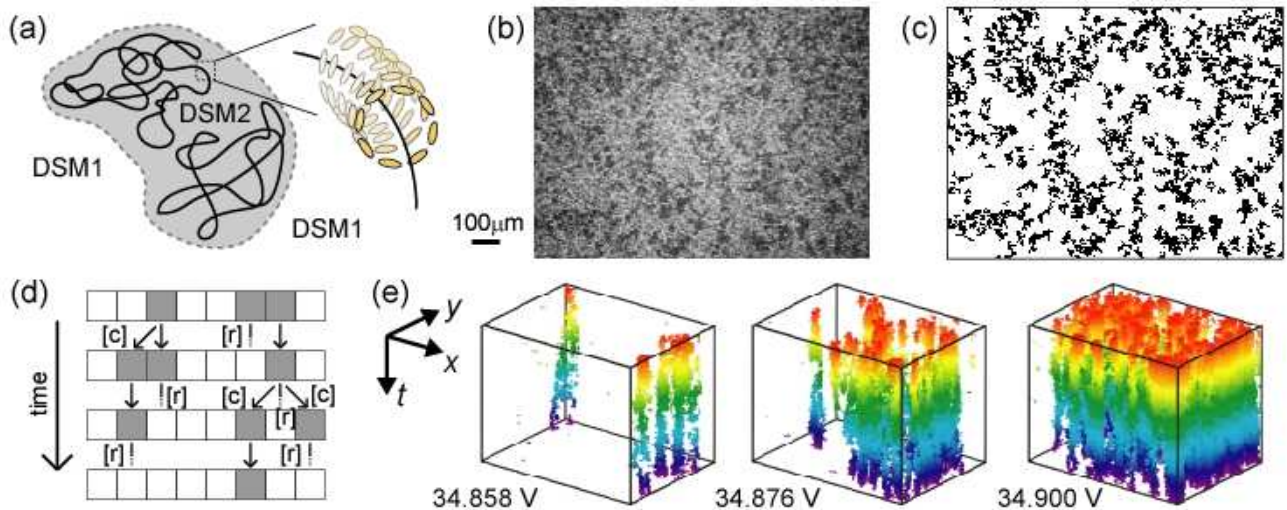}
  \caption{(Color online) Spatiotemporal intermittency between DSM1 and DSM2. (a) Sketch of a DSM2 with its many entangled disclinations, i.e. loops of singularities in orientations of liquid crystals. (b) Snapshot taken at $35.153\unit{V}$. Active (DSM2) regions appear darker than the absorbing DSM1 background. See also Movie S1 \cite{EPAPS}. (c) Binarized image of (b). See also Movie S2 \cite{EPAPS}. (d) Sketch of the dynamics: DSM2 domains (gray) stochastically contaminate [c] neighboring DSM1 regions (white) and/or relax [r] into the DSM1 state, but do not nucleate spontaneously within DSM1 regions (DSM1 is absorbing). (e) Spatiotemporal binarized diagrams showing DSM2 regions for three voltages near the critical point, namely $34.858$, $34.876$, and $34.900\unit{V}$. The diagrams are shown in the range of $1206\unit{\mu{}m} \times 899\unit{\mu{}m}$ (the whole observation area) in space and $6.6\unit{sec}$ in time.}
  \label{fig1}
 \end{center}
\end{figure*}

We chose to work within electrohydrodynamic (EHD) convection regimes,
 which occur when a thin layer of nematic liquid crystals is subjected
 to an external voltage strong enough to trigger
 the Carr-Helfrich instability \cite{deGennes_Prost-Book1993}.
One advantage of this system is that very large aspect ratios
 can easily be realized,
 and that typical timescales are small (of the order of ten milliseconds).
We focused on a transition
 between two topologically different turbulent states,
 called dynamic scattering modes 1 and 2 (DSM1 and DSM2),
 observed successively as $V$, the amplitude of the voltage, is increased
 \cite{Kai_etal-JPSJ_PRL,deGennes_Prost-Book1993}.
The difference between the two states lies in their density
 of topological defects in the director field (Fig.~\ref{fig1}a).
In the DSM2 state, these defects, called disclinations,
 are created and elongated considerably due to shear, 
 leading to the loss of macroscopic nematic order
 and to a lower light transmittance.
In DSM1, they are hardly elongated and their density remains very low.

Our quasi two-dimensional cell is made of two parallel 
 glass plates separated
 by a polyester film spacer of thickness $d= 12\,\mathrm{\mu m}$.
The inner surfaces are covered with transparent electrodes
 of size $14\,\mathrm{mm} \times 14\,\mathrm{mm}$,
 coated with polyvinyl alcohol and then rubbed
 in order to orient the liquid crystals planarly in the $x$ direction.
The cell is filled with $N$-(4-methoxybenzylidene)-4-butylaniline (MBBA)
 doped with 0.01 wt.\% of tetra-$n$-butylammonium bromide,
 maintained at temperature $25 \,\mathrm{^\circ C}$
 with a standard deviation of $2 \times 10^{-4} \,\mathrm{^\circ C}$,
 and illuminated by a stabilized light source made of white LEDs.
A CCD camera records the light transmitted through the plates.
The observed region is a central rectangle of size
 $1217 \,\mathrm{\mu m} \times 911 \,\mathrm{\mu m}$.
Since there is a minimum linear size of DSM2 domains,
 namely $d/\sqrt{2}$ \cite{Kai_etal-JPSJ_PRL},
 we can roughly estimate the number of degrees of freedom to be
 $1650 \times 1650$ for the convection area
 and $143 \times 107$ for the observation area. 
Note that the meaningful figure is that of the total
 system size, which is at least four orders of magnitude larger
 than in earlier experimental studies \cite{DPexperiments}.
In the following, we vary $V$ and fix the frequency at $250\,\mathrm{Hz}$,
roughly one third of the cutoff frequency $820 \pm 70\,\mathrm{Hz}$.

Closely above the threshold $V_{\rm c}$ marking the appearance of DSM2, 
 a regime of spatiotemporal intermittency (STI) is observed,
 with DSM2 patches moving around in a DSM1 background
 (Fig.~\ref{fig1}b and Movie S1 \cite{EPAPS}).
This suggests an absorbing phase transition \cite{Pomeau-PhysD1986}
 with DSM1 playing the role of the absorbing state (Fig.~\ref{fig1}d).
Prior to any analysis, we must distinguish DSM2 domains
 from DSM1.
This binary reduction can be easily performed by our eyes, 
 so we automated it using the facts that
 DSM2 regions look darker, have longer time correlation,
 and have minimum area of $d^2/2$ \cite{Kai_etal-JPSJ_PRL}
 (see Ref.\ \cite{EPAPS} for details).
A typical result is shown in Fig.~\ref{fig1}c and Movie S2 \cite{EPAPS}.
Figure \ref{fig1}e shows spatiotemporal diagrams 
 obtained by that means in the steady STI state.
They illustrate that the key ingredient
 is present: 
 active DSM2 patches evolve in space-time essentially 
 by contamination of the DSM1 absorbing state
 and recession from the DSM2 state (as sketched in Fig.~\ref{fig1}d).
In this context, the order parameter $\rho$ is just
 the ratio of the surface occupied by active (DSM2) regions to the whole area. 

\begin{figure}[t]
 \begin{center}
  \includegraphics[width=64.269 mm,clip]{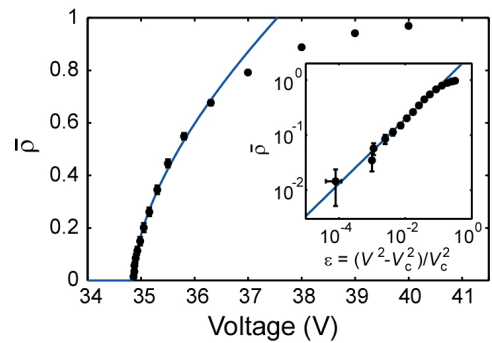}
  \caption{(Color online) Variation of the average DSM2 fraction $\bar\rho$ with $V$ in the steady state. Inset: same data in logarithmic scales, with vertical errorbars indicating the standard deviation of the time series $\rho(t)$. Fluctuations of $V$ are negligible except for the first data point, where the standard deviation is shown by a horizontal errorbar. Blue lines are fitting curves.}
  \label{fig2}
 \end{center}
\end{figure}%

\begin{figure*}[t]
 \begin{center}
  \includegraphics[width=169.654 mm,clip]{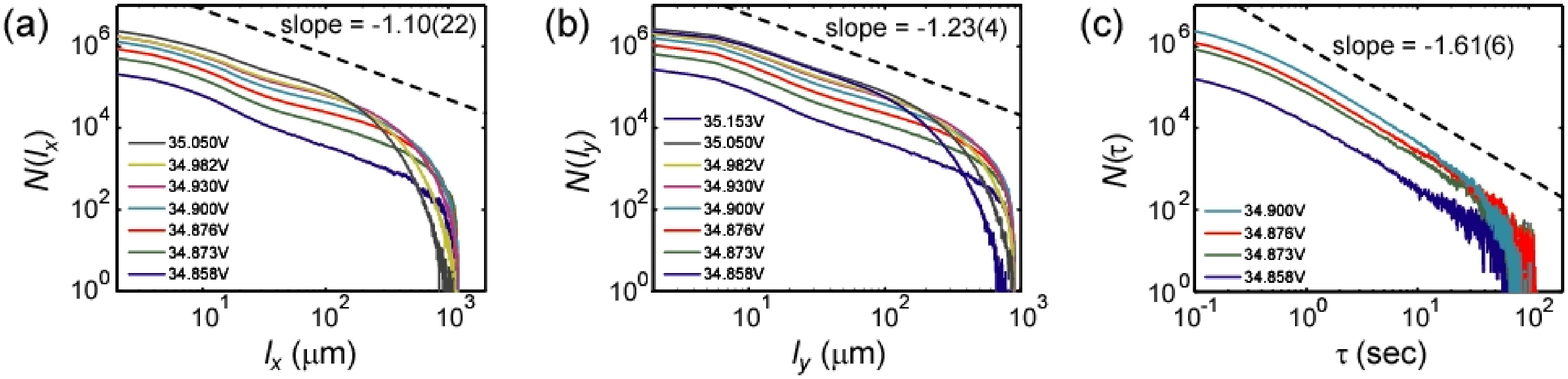}
  \caption{(Color online) Histograms of inactive (DSM1) length $l_x, l_y$ and duration $\tau$ in the steady state near criticality. Dashed lines indicate the estimated algebraic decay at threshold.}
  \label{fig3}
 \end{center}
\end{figure*}%


We first observe the steady-state STI regime under constant voltages $V$,
 in the range of $34.858\unit{V} \leq V \leq 39.998\unit{V}$.
The voltage for the onset of steady roll convection 
 is $V^* = 8.95\unit{V}$.
The time-average of $\rho$ in the steady state, $\bar\rho$,
 is taken over the period $1000 \,\mathrm{sec} < T < 8000 \,\mathrm{sec}$,
 which is longer than 1000 correlation times.
It shows a continuous and algebraic decay to zero with decreasing $V$,
 a clear signature of criticality (Fig.~\ref{fig2}).
The critical voltage $V_{\rm c}$ and the critical exponent $\beta$
 are determined by fitting these data to the usual form
 $\bar\rho \sim (V^2-V_{\rm c}^2)^{\beta}$.
(Here, deviations from criticality are measured in terms of $V^2$
 instead of $V$ by convention, since the dielectric torque that drives
 the convection is proportional to $V^2$ \cite{deGennes_Prost-Book1993}.)
This yields the following estimates
\begin{equation} 
 V_{\rm c} = 34.856 (4)\,\mathrm{V}, ~~~~\beta = 0.59(4),  \label{eq1}
\end{equation}
 where the numbers in parentheses indicate
 the uncertainty in the last digits,
 which correspond here to a 95\% confidence interval
 in the sense of Student's t.
Our estimate $\beta = 0.59(4)$ is
 in good agreement with the (2+1)-dimensional DP value
 $\beta^\mathrm{DP} = 0.583(3)$ \cite{DPexponents}.

We then measure $N(l)$ and $N(\tau)$, the distributions
 of the sizes $l$ and durations $\tau$ of the inactive (DSM1) regions.
For instance the distribution $N(l_x)$ in the $x$ direction
 is obtained by detecting all inactive segments
 in the $x$ direction for all $y$ and $t$.
We find that they decay algebraically at criticality up to a
 finite-size exponential cutoff (Figs.~\ref{fig3}a-c).
The power-law decay is fitted as
 $N(l) \sim l^{-\mu_\perp}$ and $N(\tau) \sim \tau^{-\mu_\parallel}$
 with
\begin{equation} 
 \mu_x = 1.10(22),~~~~\mu_y=1.23(4),~~~~\mu_\parallel=1.61(6),  \label{eq3}
\end{equation}
 where $\mu_x$ and $\mu_y$ indicate the exponent $\mu_\perp$
 measured in the $x$ and $y$ direction, respectively.
They are again in good agreement with the
 DP values $\mu_{\perp}^{\mathrm{DP}} = 1.204(2)$ and
 $\mu_{\parallel}^{\mathrm{DP}} = 1.5495(10)$
\cite{DPexponents}.
These exponents
 are related to the more conventional correlation length and time exponents
 $\nu_\perp$ and $\nu_\parallel$ via
 $\mu_\perp = 2-\beta/\nu_\perp$ and $\mu_\parallel = 2-\beta/\nu_\parallel$,
 which lead us to estimate
\begin{equation} 
 \nu_x = 0.66(17),~~~~\nu_y = 0.77(7),~~~~\nu_\parallel = 1.51(25). \label{eq4}
\end{equation}
They should be compared to the DP values
 $\nu_{\perp}^{\mathrm{DP}}=0.733(3)$ and 
 $\nu_{\parallel}^{\mathrm{DP}}=1.295(6)$ \cite{DPexponents}.
The above estimates for $\nu_\perp$ are
 in reasonably good agreement with the DP value,
 while the agreement on $\nu_\parallel$ is less satisfactory.
Our second set of experiments provides another,
 independent estimate of $\nu_\parallel$. 

\begin{figure}[t]
 \begin{center}
  \includegraphics[width=72.548 mm,clip]{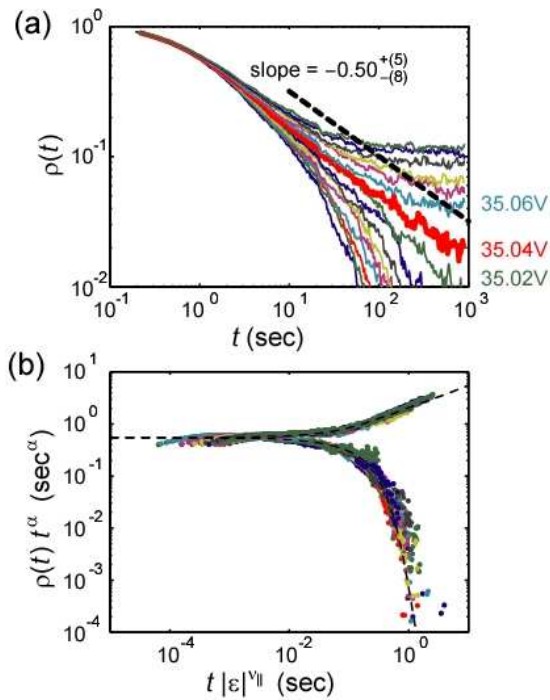}
  \caption{(Color online) Critical-quench experiments. (a) Decay of $\rho(t)$ after the quench, for $V=34.86\unit{V}$, $34.88\unit{V}$, $\cdots$, $35.16\unit{V}$ from the bottom left to the top right. The data for $V=35.04\,\mathrm{V}$ (showing the longest scaling) are indicated by a thicker line. (b) Scaling plot of data in (a),
 with $V_{\rm c}$, $\alpha$ and $\nu_\parallel$ values estimated from the experiment [Eqs.\ \eqref{eq6} and \eqref{eq7}]. The dashed curve indicates the DP universal scaling function $f(\zeta)$ obtained numerically from the process sketched in Fig.~\ref{fig1}d (so-called contact process). A collapse of similar quality is obtained when using DP-class exponent values.}
  \label{fig4}
 \end{center}
\end{figure}

Setting the voltage to $60\unit{V}$, i.e. much higher than $V_{\rm c}$,
 we wait until the system is totally invaded by DSM2 domains.
We then suddenly decrease $V$
 to a value in the range of $34.86\,\mathrm{V} \leq V \leq 35.16\,\mathrm{V}$, 
 i.e. near $V_{\rm c}$, and
 observe the time decay of activity for 900 seconds. 
We repeat this 10 times for each $V$ value and average the results 
 over this ensemble.
In such ``critical-quench experiments,'' 
 correlation lengths and times grow in time, 
 and, as long as they remain much smaller than the system size, the scaling
 estimates are free from finite-size effects.
As expected, $\rho(t)$ decays exponentially 
 with a certain correlation time for $V\le 35.02\mathrm{V}$, 
 algebraically over the whole observation time for $V=35.04\mathrm{V}$,
 and converges to a finite value for $V\ge 35.06\mathrm{V}$
 (Fig.~\ref{fig4}a).
A simple scaling Ansatz implies the following functional form
 for $\rho(t)$ in this case:
\begin{equation}  
 \rho(t) \sim t^{-\alpha} f(\varepsilon \, t^{1/\nu_\parallel}),~~~~\alpha = \beta/\nu_\parallel,  \label{eq5}
\end{equation}
 where $\varepsilon = (V^2-V_{\rm c}^2)/V_{\rm c}^2$
 is the deviation from criticality
 and $f(\zeta)$ is a universal scaling function.
From the slopes of the algebraic regimes for the three $V$
 values closest to the threshold, we estimate
\begin{equation}  
 V_{\rm c} = 35.04(2) \,\mathrm{V},~~~~\alpha = 0.50^{+(8)}_{-(5)}. \label{eq6}
\end{equation}
Note that $V_{\rm c}$ is slightly higher than in the 
 steady-state experiments. In fact,
 the roll convection onset $V^* = 8.96\,\mathrm{V}$ is
 also higher. We believe this is because, during the few days
 which separated the two sets of experiments, 
 our sample has aged, a well-known property of MBBA.
On the other hand, no measurable shift was detected during a given
 set of experiments. Our $\alpha$ value is again in
 good agreement with the DP value $\alpha^\mathrm{DP} = 0.4505(10)$
 \cite{DPexponents}.
Finally, our direct estimates of $\beta$ and $\alpha$
 in Eqs.\ \eqref{eq1} and \eqref{eq6},
 and the scaling relation \eqref{eq5} yield
\begin{equation}  
 \nu_\parallel = 1.18^{+(14)}_{-(21)},  \label{eq7}
\end{equation}
 which is in good agreement with the DP value 
 $\nu_{\parallel}^{\mathrm{DP}}=1.295(6)$ \cite{DPexponents}.
Note that our two estimates of $\nu_\parallel$ lie on opposite sides of the DP
 value, with their confidence intervals overlapping this reference number. 
Furthermore, Eq.\ \eqref{eq5} implies that
 the time series $\rho(t)$ for different voltages
 collapse on the universal function $f(\zeta)$
 when  $\rho(t)t^\alpha$ is plotted as a function of 
 $t |\varepsilon|^{\nu_\parallel}$.
Our data do collapse reasonably well
 on the universal function of DP (Fig.~\ref{fig4}b).

Continuous absorbing phase transitions are characterized by three independent
 (static) exponents \cite{Hinrichsen-AdvPhys2000}.
So far, we have found those three independent algebraic scaling laws
 (typically over two decades).
The critical exponent values all agree within a few percent
 with those of the DP class in (2+1) dimensions.
In addition, data collapse is satisfactorily achieved
 onto the universal scaling.
This constitutes the first complete and unambiguous experimental realization
 of a DP-class absorbing phase transition \cite{Takeuchi-hysteresis}.
Looking back at previous attempts
 to exhibit a DP phase transition experimentally,
 one may wonder why we obtained such clear DP scaling laws
 compared with earlier experiments (see Table S1 \cite{EPAPS}).
A major reason, already mentioned, is that
 EHD convection offers
 effective sizes orders of magnitude larger than in previously studied systems
 such as Rayleigh-B\'enard convection.
Similarly, typical timescales are much shorter.
But beyond those obvious facts,
 we believe that our unusual choice of a transition
 between two fluctuating states has also helped: in many past experiments,
 the absorbing state was essentially fluctuation-free (laminar)
 \cite{DPexperiments},
 and thus may have caused long-range effects (mean flows, rigidity of the 
laminar pattern), which probably further reduced the effective size 
of the systems and could even break DP universality 
\cite{Hinrichsen-AdvPhys2000}.
In our system, such long-range interactions are likely to be
 killed by local turbulent fluctuations of the absorbing state.
In short, our experimental setup easily reveals DP scaling
 because it is able to tame both quenched disorder (being macroscopic)
 and long-range interactions (being turbulent).

The authors are grateful to S. Kai and S. Sasa for useful discussions.
We also thank N. Oikawa, Y. Murayama, and S. Tatsumi
 for their advice on experiments.
This works is partially supported
 by Japanese Grant-in-Aid for Scientific Research
 from MEXT (1620620, 18068005)
 and by JSPS Research Fellowships
 for Young Scientists.


\begin{thebibliography}{99}

\bibitem{Hinrichsen-AdvPhys2000}
 H. Hinrichsen, Adv. Phys. \textbf{49}, 815 (2000).

\bibitem{Janssen-ZPhysB1981}
 H. K. Janssen, Z. Phys. B \textbf{42}, 151 (1981).

\bibitem{Grassberger-ZPhysB1982}
 P. Grassberger, Z. Phys. B \textbf{47}, 365 (1982).

\bibitem{Ahlers_Pikovsky-PRL2002}
 V. Ahlers and A. Pikovsky, Phys. Rev. Lett. \textbf{88}, 254101 (2002).

\bibitem{Dickman_etal-BrazJPhys2000}
 R. Dickman, M. A. Mu\~noz, A. Vespignani, and S. Zapperi, Braz. J. Phys. \textbf{30}, 27 (2000).


\bibitem{Ginelli_etal-PRE2003}
 F. Ginelli \textit{et al.}, Phys. Rev. E \textbf{68}, 065102(R) (2003).


\bibitem{DPexperiments}
 S. Ciliberto and P. Bigazzi, Phys. Rev. Lett. \textbf{60}, 286 (1988);
 F. Daviaud, M. Bonetti, and M. Dubois, Phys. Rev. A \textbf{42}, 3388 (1990);
 S. V. Buldyrev \textit{et al.}, Phys. Rev. A \textbf{45}, R8313 (1992).
 S. Michalland, M. Rabaud, and Y. Couder, Europhys. Lett. \textbf{22}, 17 (1993);
 H. Willaime, O. Cardoso, and P. Tabeling, Phys. Rev. E \textbf{48}, 288 (1993);
 M. M. Degen, I. Mutabazi, and C. D. Andereck, Phys. Rev. E \textbf{53}, 3495 (1996);
 P. W. Colovas and C. D. Andereck, Phys. Rev. E \textbf{55}, 2736 (1997);
 H. T\'{e}phany, J. Nahmias, and J. A. M. S. Duarte, Physica A \textbf{242}, 57 (1997);
 A. Daerr and S. Douady, Nature \textbf{399}, 241 (1999);
 P. Rupp, R. Richter, and I. Rehberg, Phys. Rev. E \textbf{67}, 036209 (2003);
 C. Pirat, A. Naso, J. -L. Meunier, P. Ma\"{i}ssa, and C. Mathis, Phys. Rev. Lett. \textbf{94}, 134502 (2005).

\bibitem{EPAPS}
 See EPAPS Document No. [number will be inserted by publisher] for the binarization algorithm, Movies S1 and S2, and Table S1. For more information on EPAPS, see \verb|http://www.aip.org/pubservs/epaps.html|.

\bibitem{DPopenproblem}
 P. Grassberger, in \textit{Nonlinearities in Complex Systems}, edited by S. Puri and S. Dattagupta (Narosa, New Delhi, 1997), p. 61;
H. Hinrichsen, Braz. J. Phys. \textbf{30}, 69 (2000).

\bibitem{Hooyberghs_etal-PRL2003}
 J. Hooyberghs, F. Igl\'oi, and C. Vanderzande, Phys. Rev. Lett. \textbf{90}, 100601 (2003).

\bibitem{deGennes_Prost-Book1993}
 P. G. de Gennes and J. Prost, \textit{The Physics of Liquid Crystals} (Oxford Univ. Press, Oxford, ed. 2, 1993).

\bibitem{Kai_etal-JPSJ_PRL}
 S. Kai, W. Zimmermann, M. Andoh, and N. Chizumi, J. Phys. Soc. Jpn. \textbf{58}, 3449 (1989); Phys. Rev. Lett. \textbf{64}, 1111 (1990).

\bibitem{Pomeau-PhysD1986}
 Y. Pomeau, Physica D \textbf{23}, 3 (1986).

\bibitem{DPexponents}
 P. Grassberger and Y. C. Zhang, Physica A \textbf{224}, 169 (1996);
 C. A. Voigt and R. M. Ziff, Phys. Rev. E \textbf{56}, R6241 (1997);
 for a useful list of the exponents, see
 M. A. Mu\~noz, R. Dickman, A. Vespignani, and S. Zapperi, Phys. Rev. E \textbf{59}, 6175 (1999).

\bibitem{Takeuchi-hysteresis}
Kai \textit{et al.} reported hysteresis of the DSM1-DSM2 transition
 showing algebraic dependence on the ramp rate of $V$
 \cite{Kai_etal-JPSJ_PRL}.
This may seem in contradiction with the DP framework,
 but this scaling of hysteresis loops turns out to be one aspect
 of DP criticality under the presence of a tiny residual probability
 for spontaneous nucleation of DSM2 patches.
This is detailed in K. A. Takeuchi, arXiv:0706.4152.


\end{thebibliography}
\end{document}